\documentclass[
  ]
  {aipproc}
\layoutstyle{8x11double}
\SetInternalRegister\hbadness{8000} 
\newcommand\doingARLO[2][]{%
  \ifx\mmref\undefined #1\else #2\fi
}
\begin{document}
\title{The Connection Between Spectral Evolution and GRB Lag}
\author{Dan Kocevski}{
  address={Department of Physics and Astronomy, Rice University, Houston, Texas, 77005},
  email={kocevski@rice.edu},
  thanks={Nasa}
} \iftrue
\author{Edison Liang}{
  address={Department of Physics and Astronomy, Rice University, Houston, Texas, 77005},
  email={liang@spacsun.rice.edu},
} \fi \copyrightyear  {2001}
\begin{abstract}
The observed delay in the arrival times between high and low
energy photons in gamma-ray bursts (GRBs) has been shown by Norris
et al. to be correlated to the absolute luminosity of a GRB.
Despite the apparent importance of this spectral lag, there has
yet to be a full explanation of its origin. We put forth that the
lag is directly due to the evolution of the GRB spectra. In
particular, as the energy at which the GRB's $\nu F_{\nu}$ spectra
is a maximum ($E_{pk}$) decays through the four BATSE channels,
the photon flux peak in each individual channel will inevitably be
offset producing what we measure as lag.  We test this hypothesis
by measuring the rate of $E_{pk}$ decay ($\Phi_{o}$) for a sample
of clean single peaked bursts with measured lag.  We find a direct
correlation between the decay timescale and the spectral lag,
demonstrating the relationship between time delay of the low
energy photons and the decay of $E_{pk}$. This implies that the
luminosity of a GRB is directly related to the burst's rate of
spectral evolution, which we believe begins to reveal the
underlying physics behind the lag-luminosity correlation.  We
discuss several possible mechanisms that could cause the observed
evolution and its connection to the luminosity of the burst.
\end{abstract}
\date{\today}
\maketitle
\section*{Introduction}
Gamma-ray burst spectra have a well known property of evolving as
the burst proceeds.  This evolution is characterized by two
distinct features: an overall softening of the GRB spectra with
time and a delay in the arrival of low energy photons.  Although
GRBs show remarkable variety in most of their properties, such as
duration and light curve structure, the evolution of the GRB
spectra appears to be a universal trend that is observed in a
large number of bursts.

Cheng et. al. (1995) first quantified the delay, or lag, of the
low energy photons by using the cross-correlation technique to
measure the difference in arrival times between high and low
energy photon peaks. The authors used data collected by the BATSE
instrument onboard the Compton Gamma-Ray Observatory, which for
triggering purposes was typically subdivided into four broad
energy channels from 25 keV to above 300 keV, each channel
producing a different light curve for a particular GRB event.  For
example, the four energy dependant light curves for GRB 930612 are
shown in Figure 1. When applied to timing analysis the
cross-correlation function can be used to look for variable
components between similar signals, which, for example, can yield
correlation coefficients for the temporal offset between two
photon light curves.  Using this method Cheng et. al. found that
almost all of the bursts they examined showed a delay in the 25-50
keV photon arrival times, which they attributed to scattering near
the environment surrounding the GRB.  Norris et. al. (2000) later
used a similar approach of using a cross-correlation function
(CCF) method to measure the lag between the BATSE channel 3
(100-300 keV) and channel 1 (25-50 keV) light curves for all GRBs
with independently measured redshift. What they found was an
anticorrelation between the delay in the low energy photon arrival
times and the absolute luminosity of the GRB, yielding one of the
first distance indicators that could be obtained from the
gamma-ray data alone.
\begin{figure}
\includegraphics[height=.25\textheight,keepaspectratio=true]{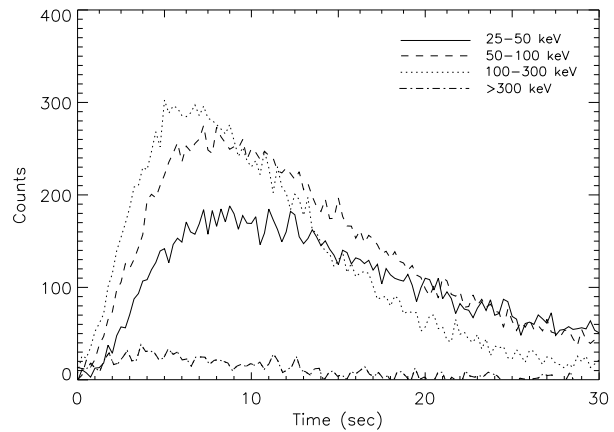}
\caption{The light curve profile of GRB 930612 (BATSE trigger
2387) in four energy bands. There is a significant time delay of
the soft energy photons.}
\end{figure}
They found that bursts with high luminosity exhibited little or no
lag, whereas fainter bursts exhibited the largest time delay.

There have been ideas proposed to explain the lag-luminosity
relationship based on viewing angles and kinematics (Solmonson
2001, Ioka 2000), but most of these do not try to address how the
lag is produced but rather why the lag is connected to the peak
luminosity.  Thus the origin of GRB lag has largely been
unexplained.

\begin{figure}
\includegraphics[height=.35\textheight,keepaspectratio=true]{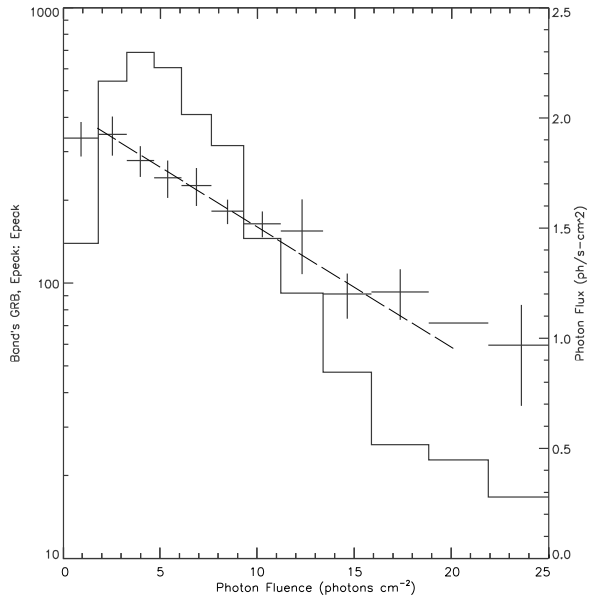}
\caption{$E_{pk}$ vs. photon fluence for GRB 990712 (7648) plotted
over the burst's photon lightcurve.  The decay constant is much
more well defined for FRED bursts.}
\end{figure}

We put forth an explanation that was first proposed quantitatively
by Brad Schaefer (in preparation astro-ph/0101462), namely that
the observed spectral lag is directly related to the evolution of
the GRB spectra to lower energies.  Therefore, as the peak in the
$\nu F_{\nu}$ spectra evolves through the various BATSE channels,
the time to peak in the individual light curves will correlate to
the hardness of $E_{pk}$.  This softening of the GRB spectra has
been known for some time (Golenetski et. al. 1983, Norris et. al.
1986) as a general hard to soft trend, but it was first quantified
by Liang and Kargatis (1996) as an exponential decay of $E_{pk}$
as a function of photon fluence,
\begin{equation}
    E_{pk} = E_{o} e^{-\Phi / \Phi_{o}}
\end{equation}
where $E_{pk}$ is the max of the $\nu F_{\nu}$ spectra, and hence
where most of the radiation energy is emitted, $\Phi(t)$ is the
photon fluence integrated from the start of the burst, and
$\Phi_{o}$ is the decay constant. In other words, the average
energy of the arriving photons becomes softer as the burst
progresses.

This simple interpretation of GRB lag predicts that the timescale
of GRB spectral decay should correlate to the burst's lag.  There
are several ways that this can be tested, but the most obvious
would be to look for a correlation between the decay constant
$\Phi_{o}$ of the $E_{pk}$ evolution and $\Delta t_{arrival}$ of
the low and high energy photons.  The decay constant represents
the e-folding rate of the break energy and therefore can be used
to parameterize the rate of evolution. It would be expected that
the bursts that have the longest decays, and hence the smallest
$\Phi_{o}$, would have the largest lag.
\begin{figure}
\includegraphics[height=.33\textheight,keepaspectratio=true]{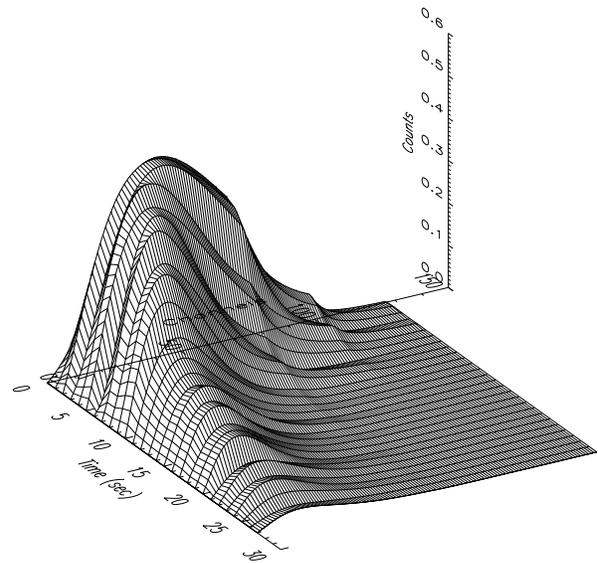}
\caption{A 3 dimensional spectral plot of GRB 911016 (907) showing
the hard to soft evolution of the peak energy. The solid line
traces the decay of $E_{pk}$.}
\end{figure}
\section*{Data Analysis}
In order to test the possible correlation between $E_{pk}$ decay
rates and spectral lag we obtained the  BATSE High Energy
Resolution (HER) data for a sample of 19 GRBs.  We then performed
time-resolved spectral fits via $\chi^2$-minimization to the
empirical Band model (Band et. al. 1993), allowing the high and
low power law indices that characterize the Band spectral model to
vary as free parameters.  These 19 bursts were chosen because they
are characterized by bright clean separable FRED (fast rise
exponential decay) pulses which tend to give reliable $\Phi_{o}$
measurements. Bursts exhibiting multiple pulse structure on short
timescales tend to have overlapping $E_{pk}$ decay periods, which
complicates the measurement of the decay constant.  For this
reason, structured bursts were excluded from this analysis. An
example of the time-resolved spectral fits that were performed is
shown in Figure 2, where the log of $E_{pk}$ vs. photon fluence is
plotted over the burst's light curve which is shown in photons
$cm^{-2}$ $s^{-1}$.  The peak energy in 990712 can be seen to
decay monotonically on the semi-log plot, and a linear fit to this
trend directly yields $\Phi_{o}$.

The lag measurements where performed by using a simple
cross-correlation (CCF) analysis between background subtracted
BATSE low (25-50 keV) and high (100-300 keV) energy light curves,
yielding the temporal offset of the two signals.  The luminosities
of these bursts were then calculated using the Norris et. al.
lag-luminosity relation, which can be expressed as (Schaefer et.
al. 2001):
\begin{equation}
    L = 2.51x10^{51} ({\Delta t_{lag}/0.1})^{-1.14}
\end{equation}
This relation gives the GRB peak luminosity as a function of
intrinsic spectral lag at the source.  Therefore, in order to
obtain meaningful luminosities we need to correct the observed lag
seen at the detector by a factor of $1+z$ for cosmological time
dilation.  Since we don't have an a priori knowledge of the
redshifts for our sample, we've employed an iteration routine that
guesses an initial value for z and then converges upon the proper
lag.  This is done in the following manner:  first an initial
guess for $z$ is used to obtain $\Delta t' = \Delta
t_{obs}/(1+z)$, which in turn gives us an initial value for the
luminosity.  This is then used with the burst's energy flux to
obtain a value for the luminosity distance $D_{L}$ to the burst.
This distance is then compared to the $D_{L}$ that can be
calculated directly from $z$ assuming standard cosmological
parameters ($H_{o} = 65$ km s$^{-1}$, $\Omega = 0.3$, $\Lambda =
0.7$).  The value for $z$ is then varied until the luminosity
distances obtained from the two separate methods converge to
within 1 part in 10$^{3}$.
\begin{figure}
\includegraphics[height=.24\textheight,keepaspectratio=true]{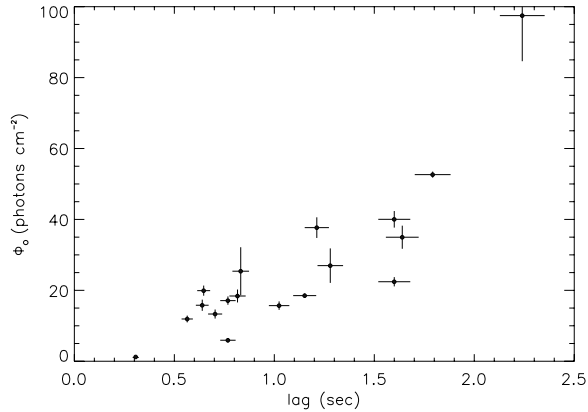}
\caption{A plot of $\Phi_{o}$ vs. the GRB spectral lag for our
entire sample.  Note that the bursts with the fastest decay (lower
$\Phi_{o}$) have the shortest lags. }
\end{figure}
It must be noted that of the seven bursts that Norris et. al. used
to find Eq. 2, only one was a FRED event.  This of course
introduces an obvious caveat in our analysis, namely that we make
the explicit assumption that the lag-luminosity relation holds for
all bursts, including single peaked FREDs.
\begin{figure}
\includegraphics[height=.24\textheight,keepaspectratio=true]{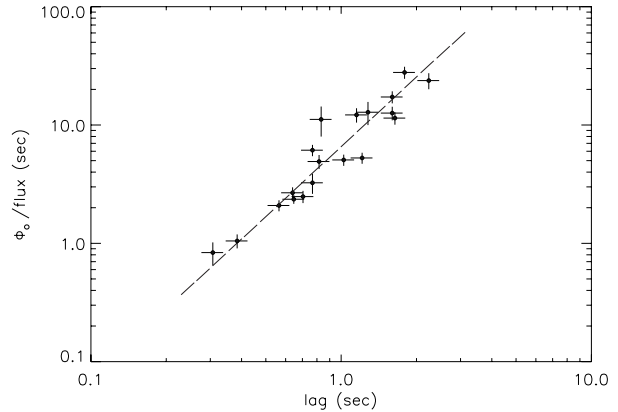}
\caption{A plot of $\Phi_{o} /$flux vs. lag for 19 GRBs}
\end{figure}
\section*{Results}
Figures 3 shows a 3 dimensional time resolved spectral plot for
GRB 911016 (trigger 907), with time on the x-axis, BATSE LAD
spectroscopic channel number on the y-axis and counts on the
z-axis.  Each slice in the yz-plane represents the GRB's time
resolved spectra, which in this case refers to a Band model fit to
the BATSE high energy resolution data, whereas a cross section
taken in the xz-plane reproduces the GRB light curve. In Figure 3,
the evolution of the peak energy, which is represented by the
solid line, can clearly be seen to begin at high energies and
decay down to the BATSE detector threshold. The individual light
curves that are used to measure the lag, typically those of
channels 1 and 3, can be thought of as being cross sections along
the xz-plane at the center of that channel's energy range.  For
example, the channel 3 light curve, which corresponds to 100-300
keV, would be located in the middle of the y-axis.  A cross
section along the xz-plane taken at this point would give a light
curve would be seen to peak very early in the burst. On the other
hand, a slice taken at the 25-50 keV energy range which
corresponds to the first few bins of the y-axis, would yield a
light curve peak very late in the burst.  We believe that this
hard to soft evolution is the fundamental factor that contributes
to the production of what we see as lag. If the decay in the above
case was very short, then the time delay between the channel 3 and
1 light curves would be relatively small, but if the decay of
$E_{pk}$ took several tens of seconds, then it's expected that the
low energy light curve would peak at a much later time. Note that
it is not necessary for $E_{pk}$ to decay through all four
channels, even if the peak energy were below 100 keV at the start
of the burst, the peak in the channel 3 light curve would still
occur very near the onset of the burst due to the geometry of the
spectra.

Figure 4 shows a plot with the resulting $\Phi_{o}$ measurement
vs. the spectral lag for our entire sample of bursts.  A general
trend can be seen that bursts with large $\Phi_{o}$ values result
in longer lags, supporting the notion that longer $E_{pk}$ decay
timescales lead to larger lags.  A power law fit to the $\Phi_{o}$
vs. lag data reveals a nearly linear correlation of the two
parameters with an index of 1.178 $\pm$ 0.058.  In Figure 5, we've
also plotted $\Phi_{o}$ normalized by the peak flux of the burst
vs. lag, which has the units of seconds, thus giving a decay
timescale in the detector frame. This seems to give a tighter
correlation than simply $\Phi_{o}$ vs. lag, and a power law fit
gives an index of 1.75. The motivations for this flux correction
comes directly from the time derivative of the Liang-Kargatis
relationship $dE_{pk}/dt = Flux/\Phi_{o}$.

\begin{figure}
\includegraphics[height=.24\textheight,keepaspectratio=true]{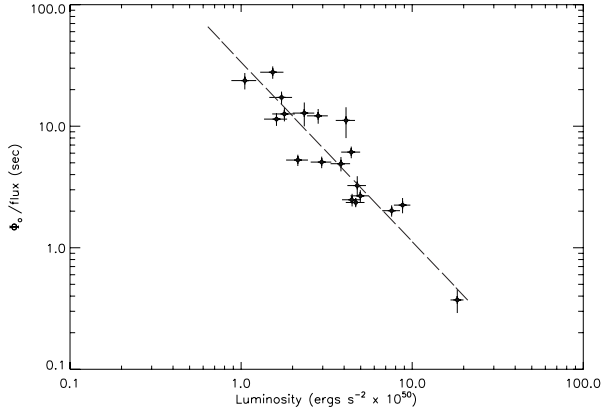}
\caption{A plot of $\Phi_{o} /$ flux vs. Luminosity as found from
the Norris et. al. lag-luminosity relation.}
\end{figure}

Because of the lag-luminosity relation, any correlation between
$\Phi_{o}$ or $\Phi_{o} /$flux vs. lag inescapably leads to a
discussion about a connection between spectral evolution and
absolute luminosity.  If the rate of spectral evolution is
linearly correlated to lag, then it should be inversely correlated
to the luminosity of the burst, assuming the validity of the
lag-luminosity relation.  To test this, we've calculated the
luminosities, and hence the distances, for all 19 bursts in our
sample using the method outlined in the previous section.  The
resulting luminosities have been plotted vs. $\Phi_{o} /$flux in
Figure 6.  Both the $\Phi_{o}$ and $\Phi_{o} /$flux relationships
satisfy the expected anticorrelation, but the $\Phi_{o} /$flux vs.
luminosity correlation provides a much tighter fit, ultimately
yielding a higher statistical significance.  These results
directly imply that more luminous bursts tend to have faster rates
of spectral evolution. It would also mean that a lower flux burst
would have a wider pulse (larger lag) compared to a high flux
burst with a similar rate of spectral decay.  This leads to two
noteworthy relationships, namely;
\begin{equation}
    L = a_{0} \bigl(\frac{flux}{\Phi_{o}}\bigr)^{a_{1}}
\end{equation}
which implies:
\begin{equation}
    D_{L} = \bigl
    [\frac{a_{0}}{\Omega}\bigl(\frac{flux^{a_{1}-1}}{\Phi_{o}^{a_{1}}}\bigr)\bigr]^{-\frac{1}{2}}
\end{equation}
where $\Omega$ is the beaming solid angle.  A best fit to the
$\Phi_{o} /$flux data gives an estimate of the $a_{0}$
coefficients to 9.8x10$^{72}$ ergs cm$^{-2}$ $\pm$ 0.39 and a
power law index of $a_{1}$ = 1.41 $\pm$ 0.06. Note that when
$a_{1}$ = 1, we recover a simple relationship between the
luminosity distance and the spectral decay constant:
\begin{equation}
    D_{L} = (\frac{a_{0}}{\Omega\Phi_{o}})^{-\frac{1}{2}}
\end{equation}
Where again, the beaming angle is left as an undetermined
parameter, as a result, the data shown in Figures 6 and 7 are
plotted per steradian.  This leaves open the possibility that any
perceived luminosity-$\Phi_{o}$ correlation may have been
distorted by the lack of beaming angle information.  The fact that
we see any correlation however, and not simply a scatter plot,
leads us to believe that the beaming angle distribution may
actually be narrow for our sample, contributing to the overall
spread in our results, but that it does not conceal the underlying
correlations.
\section*{Discussion}
The results shown in Figure 3 reveal that the spectral lag
measured in GRB is directly connected to the hard to soft
evolution of the burst spectra.  This inescapably opens up a
number of questions about the nature of the lag-luminosity
correlation.  If lag is directly proportional to the $E_{pk}$
decay constant $\Phi_{o}$, then the primary question is not why
the GRB's absolute luminosity is related to its lag, but rather
why is it related to the burst's rate of spectral evolution.  We
believe that this question is more fundamental, and ultimately
needs to be addressed.  To do so, we must first understand the
mechanism that produces breaks in the GRB spectra and what causes
its evolution. This, as it turns out, is a much harder question,
because of the uncertainties involved with the microphysics of
GRBs.  The interpretation of $E_{pk}$ (and hence its evolution)
depends on the radiation mechanism that is used to explain the GRB
spectra.  Here we attempt to review several mechanisms that could
produce the GRB specra and discuss how the hard to soft evolution
would be connected to the bursts luminosity for each model.

Liang and Kargatis originally proposed that the decay of $E_{pk}$
is governed by a confined plasma with a fixed number of particles
$N$ cooling via $\gamma$-radiation.  This type of exponential
decay of the break energy with photon fluence that is seen in the
Liang-Kargatis relation is expected if the average energy of the
emitted photons is directly proportional to the average emitting
particle energy such as in thermal bremsstrahlung or multiple
Compton scattering.  In this interpretation, the connection
between luminosity and spectral evolution arises from simple
energy conservation.  If the energy budget of a GRB pulse is
derived from a standard reservoir (Frail et. al. 2001), then more
luminous bursts are radiating their energy away faster, resulting
in a faster "cooling" of their characteristic energy, hence a
shorter lag.  In this interpretation, the quantity
$\Phi_{o}D_{L}^2$ would then represent the total number of
radiating particles.

This cooling interpretation does not work as well when applied to
the popular optically thin synchrotron model because the radiation
cooling timescales alone are typically much too short.  For
example, if we interpret the average break energy of a GRB ($\sim
E_{pk}$) as the characteristic energy of synchrotron self
absorption, then the resulting magnetic field must be extremely
high, about $\sim$ $1x10^{7}$ to $1x10^{8}$ Gauss. Separately, if
we simply assume equipartition conditions then the magnetic field
can be constrained as a function of lepton density, which, in most
models, results in a field estimate of about $10^{5}$ to $10^{6}$
G. In either case, such high $B$ fields would create a synchrotron
cooling timescale on the order of $10^{-5}$ seconds in the
comoving frame (Wu and Fenimore 2000).  This says that radiative
cooling cannot be the sole process behind the production of the
observed evolution.
\begin{figure}
\includegraphics[height=.24\textheight,keepaspectratio=true]{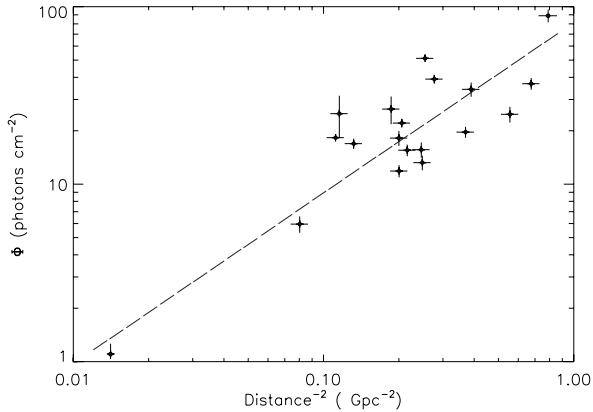}
\caption{A plot of $\Phi_{o}$ vs. distance$^{-2}$}
\end{figure}

When considering the optically thin synchrotron model, the time
variation of the burst's internal parameters must be included
namely the shell thickness $H$, $B$, the bulk Lorentz factor
$\Gamma$, and the mean number of radiating particles $N$. It is
not unlikely that these parameters will vary during the course of
the burst. In the internal shock scenario faster moving shells in
a relativistic outflow from the burst progenitor collide with
slower moving shells and emit via optically thin synchrotron
radiation. This tells us two things, that the peak emission can be
expressed as $E_{pk} \propto \gamma^{2} B$ and that $d\Gamma/dr <
0$. Therefore, as the shells expand outward, the bulk Lorentz
factor $\Gamma$ will decrease resulting in a lower observed
frequency of peak emission. This effect becomes important when
considering successive episodes of emission within the same
bursts, but the variation of $\Gamma$ is likely to be small on the
timescale of an individual pulse and hence can be considered a
constant during single periods of emission.

This leaves the size of the emitting region and the magnetic
field.  It is expected that the size of the emitting plasma will
expand in thickness once the forward and reverse shocks propagate
through the plasma. If an initial magnetic flux is frozen into the
plasma, either from the progenitor or after turbulent field growth
at the shock front, then the field strength should be inversely
proportional to the shell thickness, ($B \propto H^{-1}$),
assuming that the shells are thin. Therefore an increase in the
size of the radiating plasma will lower the magnetic field and
hence shift the synchrotron spectra to lower energies (e.g.,
Tavani 1996). Since both the forward and reverse shocks travel
close to the speed of light (i.e. $B \propto 1/ct$), this
mechanism can easily alter the spectra on the observed timescale.
Furthermore, since $E_{pk} \propto \gamma^{2} B$ and $F_{E}
\propto \gamma^{2} B^{2}$, we can recover the Liang-Kargatis
relationship (Eq. 1) in a self consistent manner, as long as the
shells can be considered thin compared to their overall size.  In
this interpretation the connection between luminosity and spectral
evolution comes directly from initial magnetic field strength and
its decay with time.  Bursts that are intrinsically more luminous
start with large magnetic fields, causing their decay timescales
to be short. This is not unlike the previous interpretation given
for the inverse Compton case, except that now the initial
variations in the magnetic field strength give rise to the
distribution in burst luminosity.

This interpretation again relies on the cooling of the radiating
leptons as a critical timescale, which is problematic since if the
magnetic field is indeed as large as that estimated earlier, then
the radiating particles would have lost all of their energy well
before the expansion of the shell takes place.  This is
representative of a larger problem with the high $B$ field
synchrotron model, namely that the cooling timescale is much
shorter then the observed variations (i.e. spikes) of emission in
GRB pulses.  Synchrotron self Compton and Synchrotron self
absorption are two models that could produce the GRB spectra with
much smaller magnetic fields and hence avoid this cooling problem
while producing evolution on the observed timescale. Furthermore,
curvature effects of a relativistically expanding shell can also
produce spectral break evolution without having to be constrained
by the burst's cooling timescale, allowing for high magnetic
fields.  A more detailed analysis of the causes behind spectral
evolution for a wider range of emission models, including the
contribution of curvature effects, will be reserved for a future
paper.

Finally, we'd like to return to the idea of a standard energy
reseviour in GRBs reported by Frail et. al 2000.  It can be shown
(Kocevski $\&$ Liang 2002) that if the energy release rate is a
constant during the prompt emission, then by means of the energy
conservation we know that if the burst is losing energy via
radiation then:
\begin{equation}
    \frac{-dE(t)}{dt} \propto L
\end{equation}
and if we apply the empirical Liang-Kargatis relation, then
\begin{equation}
    \frac{-dE_{pk}}{dt} =  \frac{F_{E}}{\Phi_{o}} =
    \frac{L}{\Omega\Phi_{o}d^2}
\end{equation}
If the total energy release is constant for GRB events, then this
would imply that the quantity $\Phi_{o}$d$^{2}$ would be a
constant for different bursts.  Kocevski $\&$ Liang (2002) have
recently reported evidence that supports this for a sample of FRED
bursts.  The constancy of $\Phi_{o}$d$^{2}$ can be also expressed
as:
\begin{equation}
    L \propto \frac{1}{\Omega}\frac{Flux}{\Phi_{o}}
\end{equation}
Which is simply the $\Phi_{o}$-luminosity anticorrelation obtained
empirically in equation 3 with $a_{1}$=1.  To test the
relationship given in Equation 6, we've plotted $\Phi_{o}$ vs.
$D_{L}^{-2}$ for our current sample of bursts in Figure 7.  From
this we can see that the data is consistent with a linear
correlation, with a best fit giving a power law index of 0.884.
Therefore, having the energy release per steradian in a GRB come
from a standard energy budget that is common among GRB events is
consistent with our results.
\begin{theacknowledgments}
We thank Brad Schaefer, Jay Salmonson, and Felix Ryde, for their
thoughtful suggestions and advice.  This work was supported by
NASA grant NAG 53824 and the NASA GSRP fellowship program.
\end{theacknowledgments}
\section*{REFERENCES}
Band, D., et al. 1993. ApJ, 428, 21\\
Cheng, L. X., et. al., 1995, A$\&$A, 300, 746\\
Crider et. al. 1997, ApJ, 479, L39\\
Golensetski, S. V. et al. 1983, Nature, 306, 451\\
Frail, D. A., et. al. 2001, ApJ, 562, L55\\
Ioka et. al. 2001 ApJ 554L, 163I\\
Liang, E., $\&$ Kargatis, V. 1996. Nature, 381, 49\\
Kocevski, D., $\&$ Liang, E., 2002, submitted, AIP Conf. Proc,
Woods Hole GRB conference\\
Norris, J. P., et. al. 1986, Apj, 301, 213\\
Norris, J. P., et al. 2000, ApJ, 534, 248\\
Salmonson, J. 2000. ApJ, 544L, 115S\\
Schaefer, B. et. al. in preparation astro-ph/0101462\\
Tavani, M., 1996, ApJ, 466, 768\\
Wu $\&$ Fenimore, 2000, Apj, 535, 29 \\
\doingARLO[\bibliographystyle{aipproc}]
          {\ifthenelse{\equal{\AIPcitestyleselect}{num}}
             {\bibliographystyle{arlonum}}
             {\bibliographystyle{arlobib}}
          }
\bibliography{sample}
\end{document}